\documentclass[twocolumn]{article}

\usepackage[pdftex]{graphicx}
\usepackage{t1enc,dfadobe}

\usepackage{cite}

\usepackage[utf8]{inputenc} 

\usepackage{hyperref}

\usepackage{xspace}
\usepackage{subfigure}

\newcommand{\FullCG}{bidirectionally linked concept-graph\xspace}

\newcommand{\CG}{BLC\xspace}
\newcommand{\eg}{\xspace{e.g.,}\xspace}

\usepackage{enumitem}
\setitemize{noitemsep,topsep=0pt,parsep=0pt,partopsep=0pt}

\title{How Sensemaking Tools Influence Display Space Usage}

\usepackage[vskip=-2pt]{quoting}

\begin{document}

\author{Thomas Geymayer$^{1}$ 
        Manuela Waldner$^{2}$ 
        Alexander Lex$^{3}$       
        Dieter Schmalstieg$^{1}$\\
         $^1$ Graz University of Technology, Austria\\
         $^2$ Vienna University of Technology, Austria \\
         $^3$ University of Utah, USA
       }

\maketitle

\begin{abstract}
We explore how the availability of a sensemaking tool influences users' knowledge externalization strategies. 
On a large display, users were asked to solve an intelligence analysis task with or without a \FullCG (\CG) to organize insights into concepts (nodes) and relations (edges). 
In \CG, both nodes and edges maintain ``deep links'' to the exact source phrases and sections in associated documents. 
In our control condition, we were able to reproduce previously described spatial organization behaviors using document windows on the large display. When using \CG, however, we found that analysts apply spatial organization to \CG nodes instead, use significantly less display space and have significantly fewer open windows.

\end{abstract}

\section{Introduction and Related Work}
\label{sec:introduction}

Extracting facts out of collections of documents (information foraging) and synthesizing information (sensemaking) are tasks widely encountered by knowledge workers.
Professional analysts employ document analysis software (``sensemaking tools''), such as the Sandbox for Analysis~\cite{wright_sandbox_2006} or Jigsaw~\cite{stasko_jigsaw:_2007}. This type of software often supports the exploration of documents through external cognition, which combines internal and external representations to perform cognitive tasks~\cite{scaife_external_1996}. Externalization of one's internal knowledge reduces the internal memory load and supports cognition by being able to directly perceive the information~\cite{kirsh_intelligent_1995}. Mind maps, concept maps, and similar visual representations of knowledge are commonly used externalization strategies, as is the spatial organization of work artifacts~\cite{malone_how_1983, kidd_marks_1994}. Indeed, Goyal et al.~\cite{goyal_effects_2013} showed that users' sensemaking performance improved significantly when provided with a visualization of shared entities across documents compared to when only provided with a note-taking tool. 

Text analytics tools, such as \textit{Jigsaw}~\cite{stasko_jigsaw:_2007} or \textit{nSpace}'s \textit{Sandbox}~\cite{wright_sandbox_2006}, categorize text entities and present them as graphs, scatter plots, or fairly free-form ``shoeboxes'', in which the connections between entities are made explicit by color coding and visual links. 
Mind mapping tools (\eg \textit{VUE}~\cite{saigal_visual_2005}) enable users not only to add text nodes, but also images and document links to the mind map for externalizing their mental concepts. 

As an alternative approach to complex sensemaking tools, large displays provide ``space to think''~\cite{andrews_space_2010}. In absence of other tools, information foraging and sensemaking is facilitated through spatial organization of information in documents and relationship extraction from multiple documents~\cite{andrews_analysts_2012}. 

Multiple researchers have found improved performance in analysis tasks when using large displays compared to small displays~\cite{czerwinski_toward_2003, reda_effects_2015}, and documented increased subjective satisfaction~\cite{andrews_space_2010, bi_comparing_2009}. Large display users employ sophisticated strategies to exploit the available space for spatial cognition, such as dividing the space into focus and context areas~\cite{grudin_partitioning_2001, bi_comparing_2009}, placing application windows as reminders~\cite{hutchings_revisiting_2004}, as well as clustering or piling windows~\cite{andrews_space_2010, waldner_display-adaptive_2011}. Large displays thereby act as \textit{externalized memory}, as users employ the space to organize and memorize information~\cite{andrews_space_2010}. In a collaborative environment~\cite{isenberg_collaborative_2009}, a large display can support the spatial arrangement and mutual awareness of opened documents.

Bradel et al.~\cite{bradel_large_2013} investigated collaborative sensemaking on a large display using either Jigsaw~\cite{stasko_jigsaw:_2007} or a simple document viewer with highlighting and annotation. They observed that users had fewer documents open with Jigsaw compared to the document viewer, but speculated that this difference was caused by the different window management behaviors of the two sensemaking tools. 
However, an alternative explanation could be that the users employed different externalization strategies in Jigsaw, so that the actual \emph{need} for multiple document windows was reduced. 

In this work, we investigate the influence of sensemaking tools on knowledge externalization strategies using a lightweight graph-based tool, the \emph{\FullCG} (\CG). 
\CG supports sensemaking with arbitrary online information sources and allows users to externalize their knowledge through a graph. It combines features of \textit{CLIP}~\cite{mahyar_supporting_2014} (attaching lists of document references to nodes and edges) and \textit{ScratchPad}~\cite{gotz_scratchpad:_2007} (referencing specific websites or passages within), and adds window layout and visual cueing capabilities that make \CG  attractive for use on large displays.
In particular, we are interested in understanding users' spatial organization strategies and \textit{whether and how users make use of the available large display space.}

Results of our user study indicate that there are strong individual differences how analysts structure their knowledge, irrespective of the sensemaking environment. However, we show that display space usage \textit{does} depend on the sensemaking environment.
From our findings, we conclude that spatial organization is a popular knowledge externalization strategy, but can happen at different scales, i.e. analysts organize abstract nodes representing facts or concepts when they have the ability to do so, and resort to organizing windows and hence make use of available display space when they do not.

\section{Study}
\label{sec:evaluation}

\begin{figure}[t!]
	\centering
	\includegraphics[width=1\linewidth]{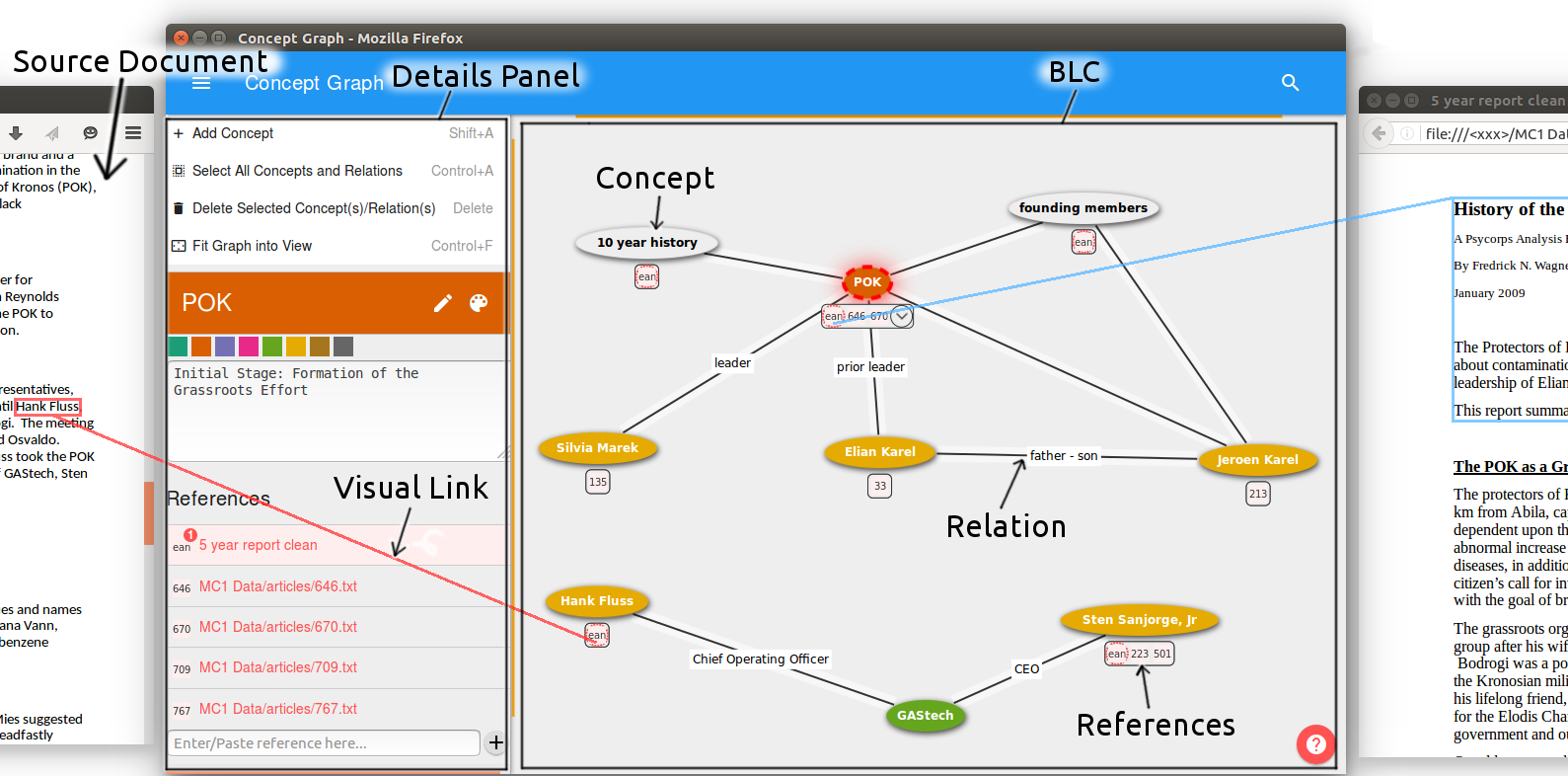}
	\vspace{-6mm}
	\caption{In \CG, concepts are rendered as nodes, relationships as links. Here, the properties of ``POK'' are shown in the detail panel on the left. References that act as evidence for a node are shown as small labels. The dashed red frame around the reference label and the red circle showing ``1'' in the reference list indicate an open document. The reference for the node ``Hank Fluss'' is opened on the left; the reference for ``POK'' on the right. Text sections that are used as evidence for nodes are highlighted and connected with visual links.}
    \label{fig:graph-with-references}
	\vspace{-4mm}
\end{figure}

We conducted a study comparing sensemaking strategies on large displays 
for an intelligence analysis task with or without \CG. 
\CG  allows users to organize browser-based information sources as mental concepts and relationships between these concepts in an interactive node-link diagram (Figure~\ref{fig:graph-with-references}).
The nodes are laid out manually, which is meant to act as memory aid and as an external representation of the user's internal knowledge~\cite{kirsh_intelligent_1995}, which can help to make inferences~\cite{larkin_why_1987}. 
Users can easily add document references to nodes and edges. 
References may refer to specific passages within these documents. 
This allows users to quickly go back to the exact piece of evidence they were previously investigating. 
When users reopen documents, windows are placed close to the referenced node. 
Hence, users can make effective use of the large display to organize their sources without considerable management overhead. 
When sources are open, visual links~\cite{geymayer_show_2014} can connect corresponding keywords between the graph and the document content. 
All these features help to organize large data sets into an ``abstracted space to think'', and expand it on demand. 

\begin{figure}[t]
	\centering
	\includegraphics[width=1\linewidth]{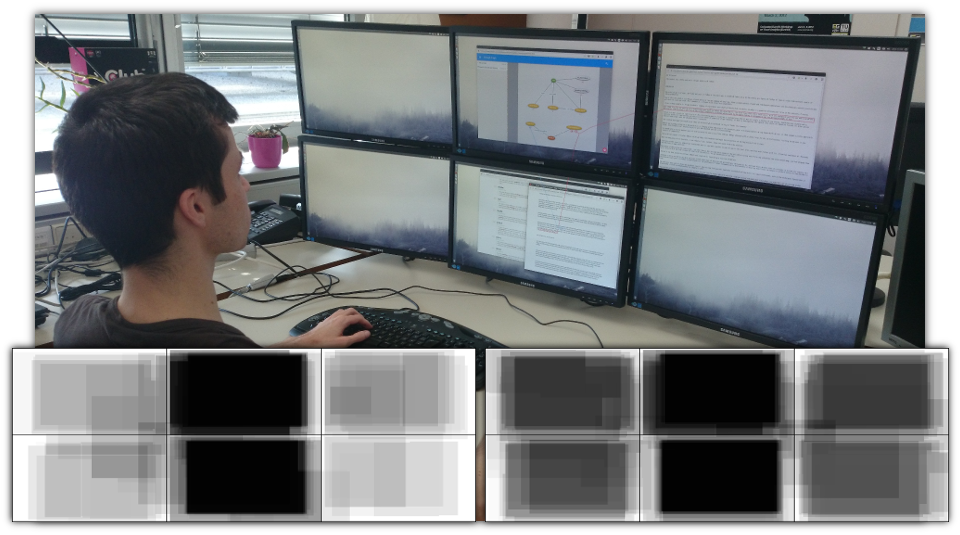}
	\vspace{-4mm}
	\caption{A user solving the sensemaking task on a large display with \CG. 
 The heat maps show the display space usages of all the users in our study in \CG (l.) and CC (r.) respectively.}
 	\label{fig:study-screen-usage}
	\vspace{-4mm}
\end{figure}

The goal of our study was to find differences in knowledge externalization strategies
between users of \CG and a condition similar to the ``Analyst's Workspace''~\cite{andrews_analysts_2012} -- a large-display sensemaking environment to support spatial organization. 
It has been shown \cite{bradel_large_2013} that users of the powerful Jigsaw analytics software rarely employ manual spatial organization. 
Our study differs, as we use a light-weight general-purpose concept graph instead of Jigsaw, and as all documents are opened in new windows in both experimental conditions. 
Our hypothesis was that --- despite having the same window management --- users of the control condition would spatially organize document windows on the large display (as observed by Andrews et al.~\cite{andrews_space_2010}), while \CG users would externalize knowledge through spatial arrangements within the concept graph. 

\subsection{Study Design}

In a between-subjects design, we assigned users to one of two conditions: 
In the control condition, \textbf{CC}, documents were shown in regular application windows with the ability to search within the corpus. In addition, visual links~\cite{geymayer_show_2014} could be invoked to connect text fragments between all open document windows. Links can be triggered from a selected word, phrase, or by searching for a term.
In contrast to Andrews et al.~\cite{andrews_analysts_2012}, we did not perform named entity extraction for visual links, to avoid introducing a confounding factor between the study conditions. 
We chose to use visual links, as they have been shown to improve performance when recognizing related items on the screen compared to simple highlighting~\cite{steinberger_context-preserving_2011} --- an aspect that is especially pronounced in large display setups. 
Users were provided with an empty Google Doc (an online text editor) to take notes. 

In the \textbf{\CG} condition, users could record information by creating nodes and edges in the graph and by adding notes to them. 
Deep linking between the concept graph and the document windows allowed for quick switching between the graph and the source information. 
Visual links could also be used for connecting arbitrary selected text between documents, as in CC, but also to connect node labels in the \CG with the web documents. No separate text document for note-taking was provided. 

\CG differs from CC in the following aspects: 
(1) it provides \emph{visual abstraction} of the contained information, (2) it organizes the information in a \emph{graph structure}, (3) it provides \emph{deep linking} between the graph nodes and edges and their associated source information, and (4) it \emph{automatically places document windows} according to the user's graph layout. 

We chose a between-subjects design, as this allowed us to use only one task, limit the length of the analysis session, and avoid learning effects. On the downside, between-subjects designs can distort the results due to individual variability. We will therefore not only report statistical significances, but also present the quantitative results visually, and provide qualitative results. 

We used the task descriptions and data from the 2011 VAST MiniChallenge 3~\cite{grinstein_vast_2011}. The data comprised around 4.500 text articles, of which 13 contained manually generated news regarding a terrorism threat in the fictitious Vastopolis area. 
The task was to identify any imminent terrorist threats in the Vastopolis metropolitan area and to provide detailed information on the threat.

\subsection{Apparatus}

The study was conducted on a PC with six 22'' monitors (1920$\times$1080) in a $3\times2$ arrangement. The user was sitting ~70cm away from the central monitor. The display setup is ~155cm wide; the displays covered a visual angle of ~95$^{\circ}$ (Figure \ref{fig:study-screen-usage}).
To search through the data, we provided users with Recoll (\url{http://www.recoll.org/}), a browser-based full-text search tool. Selecting a document in Recoll opened it in a new window with the same size as the Recoll window.
At the beginning of the session, the Recoll window was placed in the middle of the upper central monitor. For \CG, the empty \CG window was placed on the lower central monitor. For CC, an empty Google Doc was used instead.
All 20 users (10 female; age 22-49) were knowledge workers (students, researchers, or administrators). 
Sixteen users had a computer science background; all users were familiar with sensemaking tasks, such as literature research. Some users reported to have experience with dedicated tools for sensemaking or information management, such as Evernote, Mendeley, OneNote, or Trello.
After an introduction to the tools and a training session, users worked for an hour and were then asked to present their findings. After the session, they filled out a questionnaire, followed by a semi-structured interview.

\subsection{Analysis}

Sessions and interviews were recorded, and all \CG activities (concept or edge creation, adding or removing references), link activities (creation and deletion), window activities (opening, closing, moving, resizing), and query terms were logged. We compared questionnaire items, usage frequencies of the tools, display usage parameters (average/maximum number of open windows and display coverage, respectively) and the number of correctly identified plot elements in the analysis task either by Independent Samples t-tests or by Mann-Whitney U tests, if the assumption of normality was violated. 
We report significant differences, but do not mention explicitly when differences are not statistically significant. 
All interviews were transcribed and analyzed using open coding. Additionally, we qualitatively analyzed all concept graphs and Google Docs. We also analyzed all query terms and terms used for linking. 

\section{Results}



To assess the task performance of the users, we counted the number of correctly identified hints. The ground truth consists of 13 short news documents. If a reference to one of these files was added to the concept graph or the Google Doc or content of one of these files was mentioned in the interview, we counted this as correct. 

All users made an effort to follow leads and to extract a potential terrorist plot --- albeit not necessarily the correct one. 
Only few users identified elements of the ground truth plot (three \CG users and four users of CC). 
The average number of opened documents out of the set of the 13 ground truth documents was low with 1.3 in \CG and 1.6 in CC. Three \CG users and four users of CC did not open any ground truth document at all. Users' subjective satisfaction with the outcome was rated similarly for the two groups, with 3.4 in \CG and 3.0 in CC on average, on a 5-point Likert scale.

\subsection{Usage of Sensemaking Tools}

CC users conducted a significantly higher number of file queries (35.2/18.2).
However, the number of opened files was similar in \CG (31.1) as in CC (29.3). The number of \textit{distinct} files that were opened was almost equal (21.5/21.3). 

Structuring approaches were diverse across all participants, but we could observe recurring strategies. \CG users created a more or less detailed concept graph, while CC users collected text snippets and notes in the provided Google Doc. Almost all users applied some groupings on their findings. While this is inherently supported by \CG, all but one CC users also logically grouped blocks of text in their documents (\eg by source document or abstract concepts, such as ``bioterrorism'' or ``airport''). 
Additionally, users tried to maintain links to the original files. \CG users had 6-33 file references in the graph. Similarly, all but two users in CC noted file names manually in the text files. 

Many users mentioned in the interview that directly linking the source files to nodes or edges in \CG was helpful. They revisited on average 7.4 files through such references.
\CG users rated \textit{``It was very easy to find the relevant passages in the key documents again.''} significantly higher than the control group (4.5/3.7).
However, the question \textit{``I had a very good overview of the documents I had already visited''} was rated low by both groups (2.8/2.4). 
Some users criticized that the search tool Recoll did not visually mark files that have already been opened. 
\CG users explicitly noted that having numbers as file names made it hard to recall what the content of the particular file was. The node references therefore only showed up as numbers. With conventional web sources, \CG showed a favicon or, if not available, the first letter of the sources domain name.

\subsection{Display Space Usage}

\begin{figure}[t]
	\centering
	\includegraphics[width=1\linewidth]{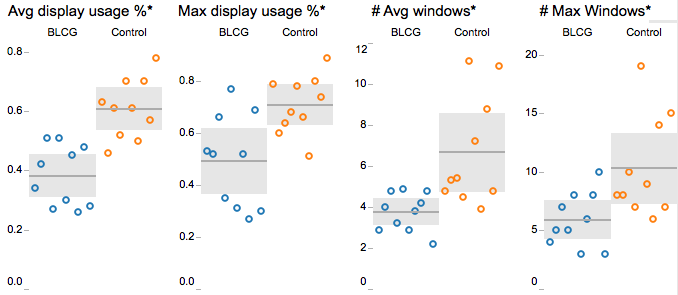}
	\vspace{-4mm}
	\caption{Usage of display space and window numbers were significantly higher in CC.}
	\label{fig:std-displays_windows}
	\vspace{-4mm}
\end{figure}

We measured display space usage as the percentage of display space covered by application windows in one-minute intervals. The average display space covered by application windows over the entire task in the control group was significantly higher than in the \CG group (61\% vs.~38\%, see Figure~\ref{fig:std-displays_windows}). This difference was similarly pronounced for the maximum display usage with 71\% control to 49\% for the \CG group, which is also statistically significant.

Figure \ref{fig:study-screen-usage} shows heat maps of how the displays were used during the analysis session. 
To create these figures, the position and size of each window was aggregated for all users. Note how, in \CG, the windows are concentrated on the two center screens, while the peripheral four monitors were used only occasionally. 

We found a considerable difference between the number of open application windows in the two conditions. The number of open application windows was significantly higher for CC than \CG (6.7/3.8 average); see Figure~\ref{fig:std-displays_windows}. The maximum number of open windows was also significantly higher for CC (10.3/5.9). Note, however, that the number of opened files (and therefore also the number of opened windows) was almost identical between those two groups (29.3 vs. 31.2)
This implies that the difference was not caused by the number of visited documents, but by the way the document windows were managed. 
Most \CG users only kept windows open occasionally. 
For instance, one user explained that \textit{``I usually closed them right after usage to keep the space tidy.''} 
CC users tended to keep documents with relevant content open for a longer time compared to \CG users, or never closed them.

We also interviewed users about display arrangement strategies. While only four \CG users mentioned a specific strategy how to arrange windows, seven users of CC were able to describe their window management strategies. We grouped interview responses into two different categories. The most popular strategy was to spatially group windows according to common concepts (described by two \CG and four CC users).
Two \CG and three CC users partitioned their display into functional units, such as a main and peripheral area. 
However, these differences are only partially reflected in the questionnaire results. 
Overall, CC group rated the large display only slightly more pleasant to use (3.7 vs.~4.3),
while, overall, the software provided was rated equally well by the two groups.

\section{Discussion and Conclusions}

We observed that users of both groups used a variety of strategies for organizing their findings and demonstrated comparable performance. 
However, users of CC had a significantly higher number of open document windows compared to \CG users. 
CC users therefore also utilized significantly more display space to spatially organize document windows. 
This implies that the \emph{outcome} of the task was hardly influenced by the provided sensemaking tools, but the \emph{process} was.

The spatial organization strategy systematically differed between the groups. Most CC users exhibited well-known spatial organization strategies with a large amount of document windows~\cite{bi_comparing_2009, andrews_space_2010,waldner_display-adaptive_2011}. Only few \CG users did that; most of them closed the text documents after reading. Instead, \CG users mainly used the graph to organize their findings. This means that both groups applied some sort of spatial organization, but it depended on the level of abstraction at which scale this spatial organization was performed. 
In fact, the concept graph can be considered a visual abstraction of the window layout, which itself is a visual externalization of the users’ mental model of the gathered information.

Thereby, we confirm a previous finding~\cite{bradel_large_2013}: Users have fewer open windows and use less space with a dedicated sensemaking tool. However, this is not exclusively caused by different window management strategies of the sensemaking environments, as suspected by Bradel et al. In our study, documents were all opened in a new window in both conditions, yet still, users handled them differently. 

We believe that the ability to easily return to the relevant passage within the original document was a major factor why \CG users reduced the number of open document windows. CC users tried to maintain a link to their original data sources by adding the file names to their text document. However, finding the relevant passages required more interaction steps than clicking on a reference and following the link in \CG. 
On the other hand, no user commented on the automatic window layout offered by the BLC.
We therefore think that this feature had a negligible effect on the users’ sensemaking strategies.

We infer that spatial organization is an important sensemaking strategy, but can happen at different scales. Meaningful visual abstractions can compress external knowledge representations. We hypothesize that the ability to easily return to the original information is a prerequisite for this compression. To formally evaluate this hypothesis, it will be necessary to compare sensemaking strategies with and without the ability to reference external documents in the future. Other important future investigations concern the scalability of the concept graph for longer activities, as well as its suitability for data with inherent chronological order or geographic reference. We expect that, due to its free-form spatial interface, \CG will also be appreciated for such kinds of tasks, but that more specialized sensemaking tools will lead to improved task performance. 

Like Bradel et al.~\cite{bradel_large_2013}, we were not able to show any measurable performance benefits (such as improved task completion time or correctness) of \CG. We speculate that this is due to the complexity of the sensemaking task, which only very few users could solve. However, our results show that users adapt their sensemaking strategies to the available environment. This is of high relevance given the heterogeneous ``device ecologies'' users are facing. 


We conclude that modern desktop managers should provide efficient means to visually abstract arbitrary information and restore it on demand. Such as feature may decrease the need to perform extensive window switching \cite{hutchings_revisiting_2004}, or to expand the display space to increase the amount of visible information. 

\section*{Acknowledgements} 
We thank Marc Streit and Markus Steinberger for helpful discussions. This work was supported in part by the Austrian Science Fund (FWF T 752-N30), and the US National Institutes of Health (U01 CA198935).


\bibliographystyle{eg-alpha-doi}
\bibliography{2016_concept-graph}

\end{document}